\documentclass[11pt,graphicx,amsmath]{article}
\usepackage{amsmath}
\usepackage{graphicx}
\usepackage{bm}

\usepackage{amssymb}

\title{ Constraints upon the spectral indices
                of relic gravitational waves by LIGO S5}

\author{\small  Y. Zhang\thanks{yzh@ustc.edu.cn}\ ,
             M.L. Tong\thanks{mltong@mail.ustc.edu.cn}\ ,
             Z.W. Fu\thanks{fuzhao@mail.ustc.edu.cn} \\
     \small Key Laboratory for Researches in Galaxies and Cosmology, CAS\\
     \small Department of Astronomy \\
     \small University of Science and Technology of China \\
     \small Hefei, Anhui, 230026,  China }
 \date{}

\begin{document}

\maketitle
\baselineskip=19truept

\def\vek{\vec{k}}
\renewcommand{\arraystretch}{1.5}
\newcommand{\be}{\begin{equation}}
\newcommand{\ee}{\end{equation}}
\newcommand{\ba}{\begin{eqnarray}}
\newcommand{\ea}{\end{eqnarray}}

\sf

\begin{center}
\Large  Abstract
\end{center}

\begin{quote}

 \sf
\baselineskip=19truept

With LIGO having achieved its design sensitivity
and the LIGO S5 strain data  being available,
constraints on the relic gravitational waves (RGWs)
becomes realistic.
The analytical spectrum of RGWs generated during inflation
depends sensitively on the initial condition,
which is generically described by
the index $\beta$,
the running index $\alpha_t$,
and the tensor-to-scalar ratio $r$.
By the LIGO S5 data of
the cross-correlated two detectors,
we obtain constraints on the parameters $(\beta, \alpha_t,r)$.
As a main result,
we have computed the theoretical signal-to noise ratio (SNR)
of RGWs for various values of  $(\beta, \alpha_t, r)$,
using the cross-correlation
for the given pair of LIGO detectors.
The constraints by  the indirect bound
on the energy density of RGWs by BBN and CMB
have been obtained,
which turn out to be still more stringent than LIGO S5.

\end{quote}


\begin{center}
{\bf 1. Introduction}
\end{center}

Recently,
LIGO S5 has experimentally obtained so far the most stringent bound
on the spectral energy density of
the stochastic background of gravitational waves,
$\Omega_0< 6.9\times 10^{-6}$ around  $\sim 100$Hz
 \cite{LIGO S5}.
Generated during inflation,
 RWGs is of cosmological origin,
and has long been investigated
\cite{grishchuk1,grishchuk,Starobinsky,Robakov},
and, in particular,
its analytical spectrum has been known \cite{zhang2}.
It depends
most sensitively upon the initial condition,
which can be generically summarized by the initial amplitude,
the spectral index $\beta$,
as well as the running index $\alpha_t$.
In particular,
small variations of $\beta$ and $\alpha_t$
will cause substantial change of the amplitude
in higher frequencies \cite{TongZhang}.
The value of $\beta$ and $\alpha_t$ are usually predicted
by specific inflationary models  \cite{Liddle}
with possible modifications by
quantum field renormalization \cite{Agullo}.
After inflation,
RGWs is  altered substantially
only by a sequence of subsequent expansions,
the reheating, the radiation era, the matter era,
and the current acceleration era \cite{zhang2},
essentially unaffected by the cosmic matter they encounter.
As a result, RGWs carry a unique information of the early Universe,
and can probe the Universe much earlier
than the cosmic microwave background (CMB).
Such cosmic processes,
as neutrino free-streaming  \cite{Weinberg,Miao},
QCD transition, and  $e^+ e^-$ annihilation \cite{WangZhang}, etc,
affect RGWs less substantially
than  small variations of $\beta$ and $\alpha_t$
around the frequency range $\sim 100$Hz of LIGO,
and can be neglected in this study.

Spreading over a broad range of frequency,
$(10^{-18}\sim 10^{10})$ Hz,
RGWs
is a major target of detectors working at various frequencies,
including  LIGO \cite{ ligo1},
Advanced LIGO \cite{ligo2},
LISA \cite{lisa},
EXPLORER \cite{Astone2007},
millisecond pulsar timing \cite{Thorsett},
and Gauss Beam \cite{fangyu}, etc.
The curl type of CMB polarization is only contributed by RGWs,
measurements of which also serve as detectors \cite{basko},
such as WMAP
\cite{Peiris,Spergel07,Komatsu,Hinshaw,Dunkley},
Planck \cite{Planck} and  CMBpol \cite{CMBpol}.
Prior to the  LIGO S5 bound \cite{LIGO S5},
often used were the  bound
from big bang nucleosynthesis (BBN)   \cite{Cyburt}
and that from the CMB anisotropy spectrum \cite{smith}.
These two indirect bounds actually  constrain
the energy density $\int \Omega_{g}(f)d\ln f$
as an integration of the spectral energy density $ \Omega_{g}(f)$.
By contrast, the LIGO S5 bound
is upon $ \Omega_{g}(f)$ itself,
and has surpassed the LIGO S4 \cite{LIGO S4}
by more than an order of magnitude.
It is now realistic
to infer from this bound some constraints
on the initial condition of RGWs
in terms of $(\beta, \alpha_t, r)$.
In this letter, using the strain data from LIGO S5 \cite{LIGO S5},
we will derive such constraints,
and compute the theoretical SNR
for the analytic spectrum of RGWs
with various $(\beta, \alpha_t, r)$.

\begin{center}
{\bf 2. Analytical Spectrum of RGWs  }
\end{center}

In a spatially flat Robertson-Walker spacetime,
the analytical mode $h_k(\tau)$ of RGWs is known \cite{zhang2}.
The spectrum at the present time $\tau_H$ is given by
\be   \label{spectrum}
h(k,\tau_H)   \equiv \frac{\sqrt{2}}{\pi}k^{3/2} |h_k(\tau_H)|,
\ee
related to
the characteristic amplitude \cite{Maggiore},
$h_c(f)= h(k,\tau_H) /\sqrt{2}$.
Here the frequency $f$ is related to
the wavenumber $k$ via $f=kH_0/2\pi \gamma$
 with $\gamma\simeq 1.97$ for $\Omega_\Lambda\simeq 0.73$ \cite{Miao}.
The spectral energy density \cite{grishchuk,TongZhang}
\be \label{omega}
 \Omega_g(f) = \frac{2\pi^2}{3}h^2_c(f)
 \left(\frac{f}{H_0} \right)^2   ,
 \ee
where $H_0 = 3.24\, h \times 10^{-18}$ Hz.
RGWs is completely fixed,
once the initial condition is given,
which is taken at the time $\tau_i$ of the horizon-crossing
during the inflation,
of a generic form \cite{TongZhang,Peiris,Komatsu}
\be \label{initialspectrum3}
h(k,\tau_i) =  \Delta_\mathcal{R}(k_0){r}^{\frac{1}{2}}
         (\frac{k}{k_0})^{2+\beta+\frac{1}{4}\alpha_t\ln{(k/k_0)}},
\ee
where the pivot wavenumber $k_0$  corresponds to a physical wavenumber
 $0.002$ Mpc$^{-1}$,
the tensor-to-scalar ratio
$r\equiv  \Delta^2_h(k_0) /\Delta^2_\mathcal{R}(k_0)$
is a re-parametrization of the normalization $\Delta_h(k_0)$
with  $\Delta^2_\mathcal{R}(k_0)=(2.445\pm0.096)\times10^{-9}$
by WMAP5+BAO+SN Mean \cite{Komatsu},
the index $\beta$ is related to the index of the power-law scale factor
during inflation $a(\tau) \propto |\tau|^{1+\beta}$ \cite{grishchuk,zhang2}
and  $\beta \simeq -2$ yields a nearly scale-invariant spectrum,
and the running index $\alpha_t$
reflects an extra bending.
Observations of CMB anisotropies
have given preliminary results on
the scalar index and the scalar running index
 \cite{Peiris,Spergel07,Komatsu}.
So far there is no observation of  $\beta$ and  $\alpha_t$,
and there are only some upper bound on $r$
 \cite{Komatsu,Hinshaw,Dunkley}.
In  scalar inflationary models,
$\beta$ and $\alpha_t$
are determined by  the inflation potential and its derivatives
  \cite{Liddle}.
There might be relations between
the tensorial indices and the scalar ones.
For generality,
we treat $(\beta, \alpha_t, r)$ as independent parameters.
In literature
the notation $n_t$ is often used,  $n_t=2\beta+4$.

\begin{center}
 {\bf 3. Constraints on the spectral indices of RGWs}
\end{center}

\begin{figure}
\centerline{\includegraphics[width=10cm]{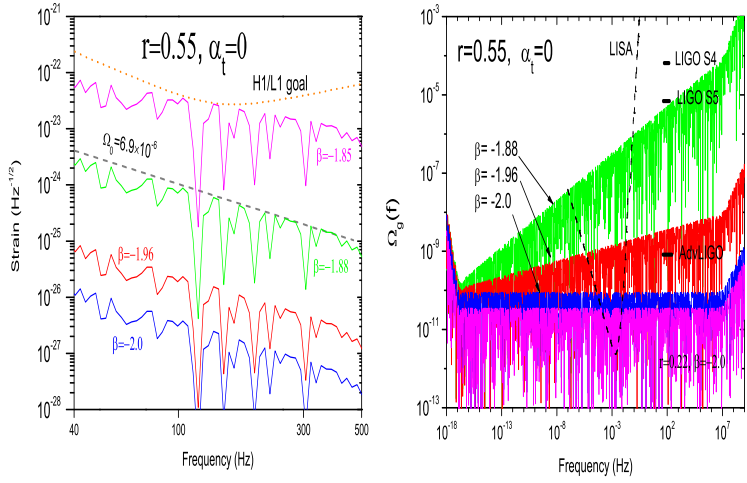}}
\caption{\label{12}
Left: The spectrum $h_c(f)\sqrt{F}/\sqrt{2 f}$
with $r=0.55$ and $\alpha_t=0$ for various $\beta$.
The dot line (labeled  by H1/L1 goal)
is the strain sensitivity $\tilde{h}_f$
of single interferometers  achieved by LIGO S5,
and the dash line ($\Omega_0=6.9\times 10^{-6}$)
is the corresponding sensitivity
from cross-correlated two interferometers
of LIGO S5 \cite{LIGO S5}.
Right: The spectral energy density
$\Omega_{g}(f)$ with $\alpha_t=0$ for various $\beta$.
The upper limit of LIGO S5
corresponds to the dash line in the left.}
\end{figure}
The left panel of Fig. \ref{12} gives the analytic spectrum
$h_c(f)\sqrt{F}/\sqrt{2 f}$ of RGWs in the frequency range $(40, 500)$ Hz
for various $\beta$ in the model $r=0.55$ and $\alpha_t=0$.
The irregular oscillations in the curves of the analytic spectra
are due to the combinations of Bessel functions
implicitly contained in the analytic solution of
RGWs \cite{zhang2,TongZhang}.
It is seen that
a small variation in $\beta$ from $-2.0$ to $-1.85$
leads to an enhancement of amplitude of $h_c(f)$
by $4$ orders of magnitude around $\sim 100$ Hz.
For  RGWs to be detectable by a single  detector
with a strain sensitivity $\tilde{h}_f$,
the condition is \cite{Maggiore},
\be\label{single}
\frac  {h_c(f)}{\sqrt{2 f}}\sqrt{F}\geq \tilde{h}_f,
\ee
where the angular factor $F=2/5$ for one  interferometer.
The dot line (labeled  by H1/L1 goal)
in the upper part of the left of Fig.\ref{12}
is the single-detector strain sensitivity
achieved by H1 and L1 of LIGO S5 \cite{LIGO S5}.
Thus, we have plotted  $h_c(f)\sqrt{F} / \sqrt{2f}$ to
directly  compare with  the strain $\tilde{h}_f$.
The single interferometers, H1 and L1,
 of the LIGO S5  put a constraint
on the index: $\beta\le -1.85$ for the model $r=0.55$ and $\alpha_t =0$.

However, by the cross-correlation of two interferometers, H1 and L1,
of the LIGO S5, the detectability  is much improved.
Approximately,
in a  narrow band  $\Delta f$ of frequencies
and a duration $T$ of observation,
the detectability condition is schematically changed to \cite{Maggiore}
\be \label{de2}
\frac  {h_c(f)}{\sqrt{2 f}}\sqrt{F}
       > \frac{1}{(2T \Delta f)^{1/4}} \tilde{h}_f,
\ee
where $ \tilde{h}_f$ is the strain of single detector.
For $T$ being long enough so that
$(2T \Delta f)^{1/4} \gg 1$,
the right hand side of Eq.(\ref{de2}) will be reduced considerably.
A detailed description of
quantitative treatment is given in Ref.\cite{Allen}.
For the case of a flat spectral energy density $\Omega_0$,
the effective strain of LIGO S5,
plotted in the dash line in left of Fig.\ref{12},
is $\sim 100$ times lower
than that from the single interferometers \cite{LIGO S5}.
This upper limit leads to
a more stringent constraint on the index:   $\beta\le -1.88$
for the same model.
This is consistent with the current observational result
 of the scalar index $n_s$
 ranging over $(0.97\sim 1.2)$ \cite{Peiris,Spergel07,Komatsu},
if a relation $n_s=2\beta+5$ is adopted,
as in scalar inflationary models.

The right of
Fig. \ref{12} gives the spectral energy density $\Omega_{g}(f)$
that corresponds to the respective spectrum $h_c(f)$
in the left.
By the upper limit  $\Omega_0=6.9\times 10^{-6}$ from
cross-correlated interferometers of LIGO S5,
the resulting constraint is $\beta\le -1.88$,
the same as from the left.
Except for the model $\beta=-2.0$ and $\alpha_t=0$,
$\Omega_{g}(f)$ is generally not flat,
and a larger $\beta$ leads to
a higher amplitude of $\Omega_g(f)$ in higher frequencies
\cite{grishchuk,TongZhang}.
$\Omega_{g}(f)$  behaves approximately
as $\Omega_{g}(f)\propto f^{0.24}$
for the model $\beta=-1.88$ and $\alpha_t=0$.
For comparison,
the sensitivity of LISA is plotted
and has a broader frequency range.

\begin{figure}
\centerline{\includegraphics[width=10cm]{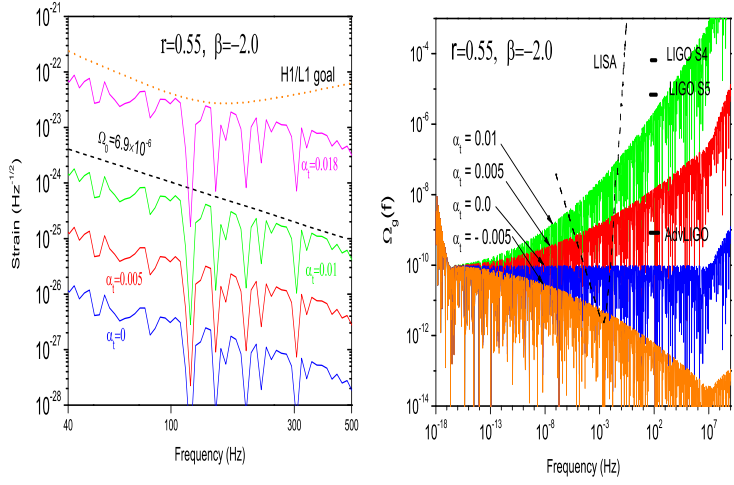}}
\caption{\label{34}
Left:  $h_c(f)\sqrt{F}/\sqrt{2 f}$ with
$r=0.55$ and $\beta=-2.0$ for various $\alpha_t$.
Right:
$\Omega_{g}(f)$ with $r=0.55$ and
$\beta=-2.0$ for various $\alpha_t$. }
\end{figure}
The left of Fig. \ref{34} plots
 $h_c(f)\sqrt{F}/\sqrt{2 f}$
for various $\alpha_t$ in the model of $r=0.55$ and $\beta=-2.0$.
A small variation in $\alpha_t$ from $0$ to $0.018$
enhances the amplitude of $h_c(f)$
by  $\sim 4$ orders of magnitude around $\sim 100$ Hz.
The single interferometers of
the LIGO S5  puts a constraint
on the running index: $\alpha_t \le 0.018$.
The cross-correlation of two interferometers of the LIGO S5
puts a more stringent constraint: $\alpha_t \le 0.01$.
So far the preliminary observed result
of the scalar running index $\alpha_s$ ranges over
$(-0.050\sim -0.077 )$ by WMAP \cite{Peiris, Spergel07,Komatsu}.
If both  RGWs and scalar perturbations are
generated by the same inflation,
one expects  $\alpha_t$ to be nearly as small as $\alpha_s$
for several kinds of smooth scalar potential \cite{Liddle}.
If so, the constraint on $\alpha_t$ by LIGO S5 is
consistent with the results  by WMAP.
The right of Fig. \ref{34} gives  $\Omega_{g}(f)$
that corresponds to those in the left.
The upper limit of
LIGO S5 gives the constraint $\alpha_t \le 0.01$,
same as that from the left.
For the model $\beta=-2.0$ and $\alpha_t=0.01$,
the slope is $\Omega_{g}(f)\propto f^{\, 0.45}$,
not flat either.

\begin{figure}
\centerline{\includegraphics[width=10cm]{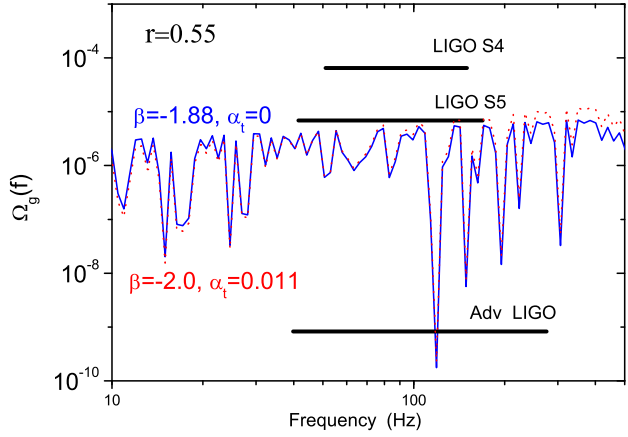}}
\caption{\label{degeneracy}
$\Omega_{g}(f)$ has the same height at $100$Hz
for the models with  $\beta= -2.0$ and $\alpha_t=0.011$,
  and with $\beta= -1.88$ and $\alpha_t=0$, respectively. }
\end{figure}

Figure~\ref{degeneracy}  shows that,
around $\sim 100$Hz,  the model $\beta=-2.0$ and $\alpha_t=0.011$
and the model $\beta=-1.88$ and $\alpha_t=0$
yield the same height of amplitude  detectable by LIGO S5.
Moreover, the slopes of $\Omega_g(f)$
in the two models only differ slightly.
Therefore,
there is a degeneracy between the indices $\beta$ and $\alpha_t$.
Given a rather narrow  frequency range, $(41.5, 169.25)$ Hz,
it is unlikely for LIGO S5 to distinguish
the spectra from these two models.
Comparatively,
LISA with a much broader frequency range
would have consequently a better chance to
distinguish models with different $\beta$ and $\alpha_t$.

The above examinations on detectability
via comparison of the spectrum $h_c(f)$ and the strain $\tilde h_f$
are still qualitative.
According to the method developed in Ref.\cite{Allen},
a more quantitative description of
the detectability is through the signal-noise ratio
\be \label{SNR}
{\rm SNR} =\frac{3H_0^2}{10\pi^2}\sqrt{T}
  \left[
  \int_{-\infty}^{\infty}df
  \frac{\gamma^2(f) \Omega_{g}^2(f)}
  {f^6 P_1(f) P_2  (f)}
  \right ]^{1/2}
\ee
for the given pair of detectors of LIGO,
where
$P_1(f)$ and $P_2(f)$ are the noise power spectrum of detector,
H1 and L1, respectively \cite{LIGO S5},
and
$\gamma(f)$ is the overlap reduction function \cite{Allen}.
Since the data of
the strain sensitivity $\tilde h_{f1}=\sqrt {P_1(f)}$
and $\tilde h_{f2}=\sqrt {P_2(f)}$ have been given  \cite{LIGO S5},
it is straightforward
to calculate SNR  from $\Omega_g(f)$ for each model.
For the model $\Omega_\Lambda=0.73$ and $\Omega_m=0.27$,
we have computed the corresponding SNR
for various indices $\beta$, and $\alpha_t$,
listed in Table 1 for $r=0.1$.
The duration $T$ in Eq.(\ref{SNR}) for LIGO S5
 is from Nov. 5, 2005 to Sep. 30, 2007 \cite{LIGO S5},
i.e.,  $T=59961600$ seconds.
Clearly, greater  values of $\beta$ and $\alpha_t$
yield higher SNR accordingly.
For other values of $r$, the corresponding SNR follows immediately
since SNR $ \propto r$.
\begin{table}
\caption{\label{table2}
The SNR for RGWs with r=0.1
for the given pair of detectors of LIGO S5
}
\begin{center}
\begin{tabular}{|l|c|c|c|c|}
\hline
& $\alpha_t=0$ & $\alpha_t=0.005$ & $\alpha_t=0.007$ & $\alpha_t=0.01$\\
\hline
$\beta=-2.0$ & $5.4\times 10^{-6}$ & $8.0\times10^{-4}$ & $6.0\times 10^{-3}$ & $1.2\times10^{-1}$\\
$\beta=-1.96$ & $2.0\times10^{-4}$ & $3.0\times10^{-2}$& $2.2\times10^{-1}$ & 4.5\\
$\beta=-1.90$ & $4.5\times 10^{-2}$ & $ 6.7$ & $5.0\times 10^1$ & $1.0\times 10^{3} $\\
$\beta=-1.88$ & $2.8\times10^{-1}$& $4.1\times10^{1}$ & $3.0\times 10^2$ & $6.2\times10^{3}$\\
\hline
\end{tabular}
\end{center}
\end{table}

\begin{center}
 {\bf 3. Constraints via the energy density $ \Omega_{gw}$}
\end{center}

Before LIGO S5 data is available in constraining
the spectrum $\Omega_{g}(f)$,
often used is the energy density parameter
\be\label{gwe}
\Omega_{gw}=
\int_{f_{low}}^{f_{upper}} \Omega_{g}(f)\frac{d f}{f},
\ee
as an integration of $\Omega_{g}(f)$ over certain frequency range,
where the cutoffs of frequencies
 depend on specific situation under consideration.
For the total energy density of RGWs in the universe,
one can take $f_{low}\simeq2\times10^{-18}$ Hz
and  $f_{upper}\simeq10^{10}$ Hz  \cite{Miao}.
Strictly speaking,
limits coming out of this method do not apply to
the spectrum $\Omega_g(f)$, and are of indirect nature.
Sometimes $\Omega_{g}(f)$ and $ \Omega_{gw}$
were used undiscriminatingly in literature.
But this will be valid only under the condition
that the integration interval $d\, f/f=d \ln f\sim 1$
and that $\Omega_{g}(f)$ be nearly frequency-independent (flat),
which is not the case for general indices $\beta$ and $\alpha_t$,
as has been demonstrated earlier.
Whenever possible,
one should distinguish  $\Omega_{g}(f)$ and $ \Omega_{gw}$
for a pertinent treatment.
Currently, two observed bounds on $ \Omega_{gw}$ are available.
One is
$\Omega_{gw}< \Omega_{BBN}\equiv1.1\times10^{-5}(N_\nu-3)$ from BBN,
where $N_\nu$ is the effective number of
relativistic species at the time of BBN.
The abundances of light-element, combined with WMAP data,
give $(N_\nu-3)<1.4$ \cite{Cyburt},
so $ \Omega_{BBN}= 1.5\times 10^{-5}$ \cite{LIGO S4}.
This bound receives contribution from frequencies down to
the lower limit $f_{low}\sim 10^{-10}$ Hz,
corresponding to the horizon scale at the time of BBN \cite{Allen}.
Another bound is
$ \Omega_{gw}<\Omega_{CMB}h^2\equiv8.4\times 10^{-6}$ at 95\% C.L.
from CMB + matter power spectrum + Ly$\alpha$
for the homogeneous initial condition of RGWs  \cite{smith}.
For  the Hubble parameter  $h=0.701$ \cite{Komatsu,Hinshaw},
this is $\Omega_{CMB} = 1.62\times 10^{-5}$,
receiving contributions from frequencies down to a much lower limit
$f_{low}\sim 10^{-15}$ Hz,
corresponding to the horizon scale at the decoupling for CMB.
From the theoretical side,
substituting the analytical spectrum $\Omega_g(f)$
as the integrand into Eq.(\ref{gwe}),
the resulting integral $\Omega_{gw}$ is a function of the indices
$\beta$ and $\alpha_t$,
since $\Omega_g(f)$  intrinsically depends on $\beta$ and $\alpha_t$.
By this way,
we can derive constraints on $\beta$ and $\alpha_t$
by the bounds $\Omega_{BBN}$ and $\Omega_{CMB}$.
In  carrying out the integration,
we take  the upper limit of integration
 $f_{upper}= 10^{10} $ Hz \cite{Miao}.
As for the lower limit,
we take  $f_{low}=10^{-10}$Hz for BBN case,
and $f_{low}=10^{-15}$Hz for CMB case, respectively.
It turns out that the  integral $\Omega_{gw}$ is  sensitive
to the value of $f_{low}$
for very small $\beta$ and $\alpha_t$.

\begin{figure}
\centerline{\includegraphics[width=10cm]{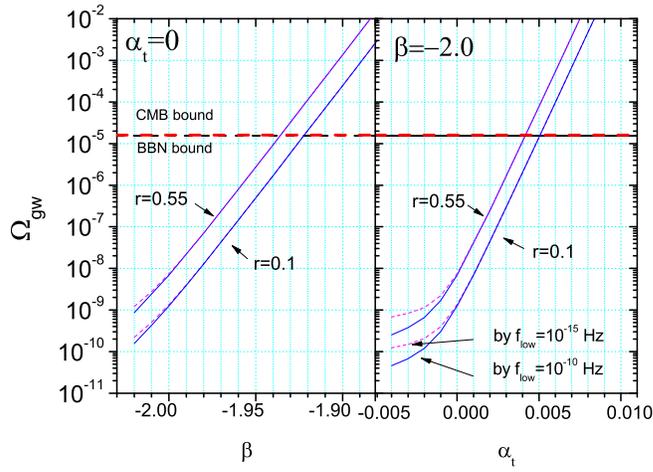}}
\caption{  \label{BBN}
Left: $\Omega_{gw}$ as a function of $\beta$.
Right: $\Omega_{gw}$ as a function of $\alpha_t$.
}
\end{figure}

The left of Fig.\ref{BBN}
shows the $\beta-$dependence of  $\Omega_{gw}$
for fixed $\alpha_t=0$ and $r=0.55$ and  $0.1$,
and the right shows the $\alpha_t-$dependence of  $\Omega_{gw}$
for fixed $\beta=-2.0$ and $r=0.55$ and  $0.1$.
In Fig.\ref{BBN} the  horizontal dash lines
are the bounds $\Omega_{BBN}$ and $\Omega_{CMB}$,
which are close to each other.
The resulting constraint on $\beta$ is
$\beta \lesssim -1.96$ for $r=0.55$ and $\alpha_t=0$,
and $\beta \lesssim -1.98$ for $r=0.1$ and $\alpha_t=0$.
The resulting constraint on $\alpha_t$ is
$\alpha_t \lesssim 0.004$ for $r=0.55$ and $\beta =-2.0$,
and $\alpha_t  \lesssim 0.005$ for $r=0.1$ and $\beta =-2.0$.
These constraints on $\beta$ and $\alpha_t$ by BBN and CMB
are more stringent than those by LIGO S5.

~
~

ACKNOWLEDGMENT:
Y. Zhang's work has
been supported by the CNSF No. 10773009, SRFDP, and CAS.
M. L. Tong's work is partially supported by Graduate
Student Research Funding from USTC.

\end{document}